\documentclass{article}
\usepackage{amsfonts,amssymb, amsmath,mathrsfs}
\usepackage[english]{babel}

\def\nn{\nonumber }
\def\bq{ \begin{equation} }
\def\eq{ \end{equation} }
\def\ben{ \begin{eqnarray} }
\def\en{ \end{eqnarray} }

\def\g{{\gamma}}

\newtheorem{prop}{Proposition}

\newtheorem{re}{Remark}
\newenvironment{rem}{\begin{re} \rm }{\end{re}}
\begin{document}


\title{One  invariant measure and different Poisson brackets for two nonholonomic systems}
\author{A.V. Tsiganov \\
\it\small
St.Petersburg State University, St.Petersburg, Russia\\
\it\small e--mail:  andrey.tsiganov@gmail.com}

\date{}
\maketitle
\begin{abstract}
We discuss  the  nonholonomic Chaplygin and the Borisov-Mamaev-Fedorov systems, for which
 symplectic forms are  different deformations  of the square root from the corresponding
 invariant volume form.  In both cases second  Poisson bivectors are determined by $L$-tensors  with non-zero torsion on the configurational space, in contrast with the well known  Eisenhart-Benenti and Turiel constructions.
 \end{abstract}

\section{Introduction}
\setcounter{equation}{0}
At the beginning of the 20th century  S.A. Chaplygin  showed that a two degree of freedom
non\-ho\-lo\-no\-mic system possessing invariant measure can be reduced to  Hamiltonian after a suitable change
of time \cite{ch03,ch11}.  Using this  process of the Chaplygin  hamiltonization   we can get  the usual Hamilton-Jacobi equation, variables of separation, the Abel-Jacobi equations, quadratures, etc \cite{bal10,bl09,bm01,bm08,bm08b,ekm04,bl11}.

In  bi-hamiltonian geometry  separability is invariant geometric property of the distribution defined by mutually commuting independent functions $H_1,\ldots, H_n$ \cite{ts07a,ts11s,ts10d}.  In fact there is  neither Hamilton-Jacobi equation, no time which describes only  some partial parametrization of geometric objects. So, in this paper we want to show how these standard bi-Hamiltonian geometric methods may be directly applied to nonholonomic systems without any change of time. The second aim is to discuss  a deformation of the Turiel construction \cite{Turiel}, which appears only  in the nonholonomic case and gives rise to  some interesting modifications  of the standard  conformal Killing tensors  that lie at the heart of  classical Eisenhart-Benenti  theory of separability \cite{ben97,ben05,tsg05}.

As an  example,  we will consider a rolling of dynamically asymmetric and balanced ball over an
 absolutely rough fixed  sphere with  radius $a$.    At $a\to \infty$  one gets  a Chaplygin
problem on a non-homogeneous sphere rolling over a horizontal plane without slipping  \cite{ch03}. Thus, we
are  able to compare  bi-Hamiltonian methods  with the  the Chaplygin  hamiltonization.

 Let $\omega=(\omega_1,\omega_2,\omega_3)$ be an angular velocity vector  of the rolling  ball.  Its  mass,  inertia tensor and radius
 will be denoted by  $m$, $\mathbf I = \mathrm{diag}(I_1, I_2, I_3 )$ and $b$, respectively.  According to  \cite{bf95}, the angular momentum  $M=(M_1,M_2,M_3)$ of the ball with respect to the contact point with the sphere  is equal to
  \bq\label{om-m}
M=(\mathbf I+d\mathbf E)\,\omega-d(\g,\omega)\g\,,\qquad \qquad d=mb^2.
\eq
Here $\g=(\g_1,\g_2,\g_3)$ is the unit normal vector to the fixed sphere at the contact point, $\mathbf E$ is the unit matrix and $(.,.)$ means the standard scalar product in $\mathbb R^3$. All these vectors are expressed in the so-called body frame,  which is firmly attached to the ball, and  its axes coincide with the principal inertia axes of the ball.

For further use we rewrite the relation (\ref{om-m})  in the following equivalent form
\bq\label{w12}
  \omega=\mathbf A_g\,M\equiv\Bigl( \mathbf A +{d}\,{\mathrm g(\g)}\,\mathbf A\,\bigl(\g \otimes\g\bigr)\,\mathbf A \Bigr) M\,,
\eq
where
\[
\mathbf A=\left(
            \begin{array}{ccc}
              a_1 & 0 & 0 \\
              0 & a_2 & 0 \\
              0 & 0 & a_3
            \end{array}
          \right)=(\mathbf I+d\mathbf E)^{-1}\,,
\]
and
\bq\label{g-fun}
\mathrm g(\g)=\dfrac{1}{1-d (\g,\mathbf A\g)}\,.
\eq
According to \cite{ch03,bf95},  there is  no slip nonholonomic constraint  associated with the zero velocity in the point of contact. It allows us to reduce equation of motion to the following form
\bq\label{m-eq}\dot M=M\times \omega\,,\qquad \dot \g=\kappa \g\times \omega\,.
\eq
where $\times$ means vector product  in $\mathbb R^3$ and  $\kappa=a/(a+b)$.

At any $\kappa$  there are  three  integrals of motion
\bq\label{3-int}
H_1=(M,\omega)\,,\qquad H_2=(M,M)\,,\qquad  C_1=(\g,\g)\,,
\eq
and invariant measure
\bq\label{rho}
\quad \mu= \sqrt{g(\g)\,}\mathrm d\g \mathrm dM\,.
\eq
If  $\kappa=\pm 1$ there is one more integral of motion
\bq\label{4-int}
C_2=(\g,\mathbf B M)\,,\qquad
\mathbf B=\left(
            \begin{array}{ccc}
              b_1 & 0 & 0 \\
              0 & b_2 & 0 \\
              0 & 0 & b_3
            \end{array}
          \right)=\mbox{tr}\, \mathbf A^{-1}+(\kappa-1)\mathbf A^{-1}\,.
\eq
At  $ d=0$ we have $ g(\g)=0$ and $\omega=\mathbf A M$. In this case $\mu$ is a standard volume form with constant density and, therefore,  equations (\ref{m-eq}) describe some hamiltonian flow.  Namely,  at  $\kappa=1$  equations (\ref{m-eq}) can be identified with the Euler-Poisson equations describing the rotation of a rigid body around a fixed point,  whereas at $\kappa=-1$   equations (\ref{m-eq}) describe the so-called  Contensou model of Fleuriais gyroscope \cite{cont}.

At  $ d\neq 0$ equations (\ref{m-eq}) describe the rolling of a dynamically nonsymmetric sphere over a  horizontal plane or a  fixed sphere without slipping. The case $\kappa=1$ is the so-called Chaplygin system  \cite{ch03} and case $\kappa=-1$ will be referred as the Borisov-Mamaev-Fedorov system. A detailed  description of these nonholonomic systems may be found in \cite{bf95, bm01,bmm08,bm08,bm08b,ekm04}.

In  \cite{bmm08,bm08,bm08b} authors  change  time variable in the equations of motion (\ref{m-eq}) first and only then study the Poisson structure of the resulting equations. In the Hamiltonian mechanics   transformation of time can drastically change almost all the invariant  geometric properties of the initial system, such as the Lagrangian foliation, compatible Poisson structures, Lax and $r$-matrices, bi-Hamiltonian construction of the variables of separation, etc \cite{ts00,ts01}.  For the nonholonomic system it can change even  the initial Hamilton function \cite{bal10,ekm04,bl11}.

Our main aim is to get  a family of  Poisson brackets associated with the invariant measure (\ref{rho}) in framework of the bi-Hamiltonian geometry, i.e. without any change of  time. It means that we for a while forget about the equations of motion (\ref{m-eq}) and try to solve  the following geometric equations
\bq\label{geom-eq}
PdC_{1,2}=0\,,\qquad (PdH_1,dH_2)\equiv\{H_1,H_2\}=0\,,\qquad [P,P]=0\,,
\eq
where $[.,.]$ is the Schouten bracket, with respect to the Poisson bivector $P$.

So, in our approach the Hamiltonization process is equivalent to a search of the Poisson structure satisfying to equations (\ref{geom-eq}), i.e.
\[\mbox{Hamiltonization}\quad\Longleftrightarrow\quad\mbox{Poisson bracket}\]
because using this bracket we can always get new Hamiltonian system
\[
\dfrac{\mathrm d}{\mathrm dt'}\,z_k=\{H_1,z_k\}\,,
\]
with new time $t'$ and the same integrals of motion in involution.

\begin{rem}
In our case at $d=0$ we have Hamiltonian system with canonical Poisson bracket. So, Hamiltonization is equivalent to existence of the proper deformations of this canonical Poisson bracket. Obstacles to such deformations are well known, see, for instance, the geometric quantization theory.
\end{rem}

The principal disadvantage is that equations  (\ref{geom-eq}) have infinitely many solutions  \cite{ts07a,ts10k,ts11s}.  So, in order to get any particular solution we have to set aside an invariance and to  narrow the search space.

\subsection{Spherical coordinates}
In order to  to  narrow the search space we will use the notion of natural Poisson bivectors on the Riemannian   manifolds \cite{ts11s}.  In this case we have to reduce our initial phase space to the cotangent bundle of the unit two-dimensional
Poisson sphere.

Namely, we can  avoid the  solution of the first  equations $PdC_{1,2}=0$  in (\ref{geom-eq})  using the slightly modified Euler variables
\bq\label{sph-coord}
\begin{array}{ll}
\g_1 =\sin\phi\sin\theta,\qquad &
M_1 =\dfrac{1}{b_1}\left(\dfrac{\sin\phi}{\sin\theta}\,\bigl(\cos\theta\,p_\phi+p_\psi\bigr)-\cos\phi\,p_\theta\right)\,,\\
\\
\g_2 = \cos\phi\sin\theta,\qquad&
M_2=\dfrac{1}{b_2}\left(\dfrac{\cos\phi}{\sin\theta}\,\bigl(\cos\theta\,p_\phi+p_\psi\bigr)+\sin\phi\,p_\theta\right)\,, \\
\\
\g_3 =\cos\theta\,,\qquad &M_3 = -\dfrac{p_\phi}{b_3}\,,
\end{array}
\eq
such as
\[
C_1=(\g,\g)=1\,,\qquad C_2=(\g,\mathbf B M)=p_\psi.
\]
In the generic case the remaining equations in (\ref{geom-eq}) have infinitely many solutions \cite{ts07a}. In order to find  at list one  particular solution  we impose the following restriction
\bq\label{restr}
C_2=(\g,\mathbf B M)=p_\psi=0\,.
\eq
In this case we have natural integrals of motion $H_{1,2}$ (\ref{3-int}) and, therefore,  we can solve our equations
using the notion of natural Poisson bivectors \cite{ts11s}.

At $\kappa=1$ coordinates $(\phi,\theta)$ in (\ref{sph-coord}) are  usual spherical coordinates on the unit sphere $S^2$  at
\[\kappa=1\,,\qquad \qquad b_1=b_2=b_3=\mbox{tr}\,\mathbf A^{-1}=1\,.\]
At $\kappa=-1$  we  replace parameters $a_i$ and $J_i=a_i^{-1}$ on $b_i$
\[\kappa=-1\,,\qquad\qquad
a_1=\dfrac{2}{b_2+b_3}\,,\qquad a_2=\dfrac{2}{b_1+b_3}\,,\qquad
a_3=\dfrac{2}{b_1+b_2}\,,
\]
in order to get  more foreseeable formulas.

After the  change of variables (\ref{sph-coord}), at $\kappa=\pm1$ we obtain two different dynamical systems on the common  phase space $\mathcal M$ which is topologically equivalent to cotangent bundle $T^*S^2$ to the sphere. These systems have a common invariant volume form (\ref{rho}) and two different Poisson structures, see next Section.

In order to show the difference between the Chaplygin and Borisov-Mamaev-Fedorov systems we present one equation of motion
 \ben
\kappa=1\,,\qquad \dot{\theta}&=&\dfrac{(a_1-a_2)\sin2\phi}{2}\left(\dfrac{\cos\theta}{\sin\theta} -
\dfrac{\mathrm g(a_3-a_1\sin^2\phi-a_2\cos^2\phi)\sin2\theta}{2}
\right)p_\phi
\nn\\ \nn\\
&-&\left(\dfrac{\mathrm g(a_1-a_2)^2\sin^22\phi\sin^2\theta}{4}+a_1\cos^2\phi+a_2\sin^2\phi\right)p_\theta\nn\\
\nn\\
\kappa=-1\,,\qquad \dot{\theta}&=&\dfrac{1}{b_1b_2(b_1+b_3)(b_2+b_3)}\left[
\left(\dfrac{b_3(b_1-b_2)\sin2\phi\cos\theta}{\sin\theta}\right.\right.\nn\\
\nn\\
&-&\left.\dfrac{\mathrm g(b_1-b_2)\bigl(b_3(b_1^2\cos^2\phi+b_2^2\sin^2\phi)-b_1^2b_2^2\bigr) \sin2\phi\sin2\theta}{b_3(b_1+b_2)(b_1+b_3)(b_2+b_3)}\right)\,p_\phi\nn\\
\nn\\
&+&\left.\left(2\bigl(b_3(b_1\sin^2\phi+ b_2\cos^2\phi)+b_1b_2\bigr)-\dfrac{\mathrm g b_3(b_1-b_2)^2}{(b_1+b_3)(b_2+b_3)\sin^22\phi\sin^2\theta}\right)p_\theta\right].
\nn
\en
and one  integral of motion
\[\begin{array}{ll}
\kappa=1\,,\qquad & H_2=\dfrac{1}{\sin^2\,\theta}\,p_\phi^2+p_\theta^2\,,\\
\\
\kappa=-1\,,\qquad& H_2=\left(\dfrac{(b_1^2\cos^2\phi+b_2^2\sin^2\phi)\cos^2\theta}{b_1^2b_2^2\sin^2\theta}+\dfrac{1}{b_3^2}\right)p_\phi^2
+\dfrac{(b_1^2-b_2^2)\sin2\phi\cos\theta}{b_1^2b_2^2\sin\theta}\,p_\phi\,p_\theta\\
\\
\qquad&+\dfrac{b_1^2\sin^2\phi+b_2^2\cos^2\phi}{b_1^2b_2^2}\,p_\theta^2\,.
\end{array}
\]
Of course, any calculations for the Borisov-Mamaev-Fedorov systems require  more  efforts and large-scale resources  in comparison to the same calculations for the Chaplygin system.

\section{Invariant measure and Poisson brackets}
\setcounter{equation}{0}

Let $\mathcal M$  be   a smooth symplectic manifold endowed with a symplectic form  $\Omega$ which in the Darboux coordinates
\[z=(q,p)=(q_1,\dots,q_n,p_1,\dots,p_n)\]
reads as
 \bq\label{omega-0}
\Omega=\mathrm dp_1\wedge dq_1+\ldots
 \mathrm dp_n\wedge dq_n\,.
 \eq
The volume form $\Omega^2$ on $\mathcal M$ is invariant under all hamiltonian diffeomorphisms by the Liouville theorem.

If we have another invariant  volume form $\mu$ on the same manifold $\mathcal M$, we can  get another symplectic form $\Omega_\mu$ taking a formal square root on $\mu$, because
\bq\label{sqrt-nu}
\mu=\Omega_\mu^2.
\eq
However, in our case  invariant volume form  $\mu=\sqrt{\mathrm g}\,\lambda$  (\ref{rho})  is invariant with respect to the non-hamiltonian flow (\ref{m-eq}) and, therefore, we have to deform its formal square root (\ref{sqrt-nu}). We will describe these deformations using  Poisson bivectors, instead of the corresponding symplectic forms.

We rewrite the  Poisson bivector $P$ associated with the canonical  symplectic form  $\Omega$ (\ref{omega-0})    in the following tensor form
\bq\label{p1}
P= \left(
 \begin{array}{cc}
 0 & L_{ij} \\
 \\
 -L_{ij}\qquad&\displaystyle \sum_{k=1}^n\left(\dfrac{\partial L_{ki}}{\partial q_j}-\dfrac{\partial L_{kj}}{\partial q_i}\right)p_k
 \end{array}
 \right)=\left(
 \begin{array}{cc}
 0 & \mathrm{Id} \\
 \\
 -\mathrm{Id} & 0
 \end{array}
\right)\,,
\eq
where $L$ is an identity (1,1)  tensor field on a configurational space.

We use such unusual notation because  any  torsionless (1,1) tensor field  $L'(q_1,\ldots,q_n)$ on a configurational space $Q$ with coordinates $q_1,\ldots,q_n$ determines another Poisson bivector
\bq \label{p2}
 P'= \left(
 \begin{array}{cc}
 0 & L'_{ij} \\
 \\
 -L'_{ij}\qquad&\displaystyle \sum_{k=1}^n\left(\dfrac{\partial L'_{ki}}{\partial q_j}-\dfrac{\partial L'_{kj}}{\partial q_i}\right)p_k
 \end{array}
 \right)\,
 \eq
 on $\mathcal M$, according to  \cite{Turiel}. The corresponding Poisson brackets read as
\[
\{q_i,q_j\}' = 0 \,,\qquad \{q_i,p_j\}'=L'_{ij}\,,\qquad
\{p_i,p_j\}'=\sum_{k=1}^n\left(\dfrac{\partial L'_{ki}}{\partial q_j}-\dfrac{\partial L'_{kj}}{\partial q_i}\right)p_k\,.
\]
The vanishing of $L'$ torsion entails that $P'$ (\ref{p2}) is a Poisson bivector compatible with $P$, i.e.
\[ [P,P]=[P',P]=[P',P']=0.\]
The torsion of the  (1,1) tensor field  $A$  equals to zero, if for any vector fields $X,Y$
\[
T_A(X,Y)\equiv\mathcal L_{AX}\,AY-A\Bigl(\mathcal L_{AX}\,Y+\mathcal L_{AY}\,X-A\mathcal L_X\,Y\Bigr)=0\,,\qquad \forall X,Y\,.
\]
 Here $\mathcal L_X$ means  the Lie derivative along $X$.

\begin{rem}
In the framework of the Eisenhart-Benenti  theory tensor field $L'$ yields  special conformal Killing tensor called the Benenti $L$-tensor,  Killing-St\"{a}ckel space, St\"{a}ckel web etc  \cite{ben97,ben05}.  Of course, we can try to transfer the corresponding geometric machinery to nonholonomic theory.
\end{rem}

One possible generalization of the Turiel  construction  (\ref{p2}) has been proposed in \cite{ts11s}. Here we consider some other generalizations related with the nonholonomic integrable systems.

In our case $n=2$ and the Darboux coordinates on $\mathcal M=T^* S^2$ are standard spherical coordinates (\ref{sph-coord}), thus,
\ben
q_1=\phi\,,\quad q_2=\theta\,,\quad p_1=p_\phi\,,\quad p_2=p_\theta\,.
\en
At $d=0$ and $\kappa=\pm1$ we have the constant invariant measure  and, therefore,
integrals of motion $H_{1,2}$ (\ref{3-int})  are in involution
\[
\{H_1,H_2\}=0\,,\qquad \qquad d=0,\qquad \kappa=\pm1\,,
\]
with respect to the Poisson brackets associated with the canonical bivector $P$ (\ref{p1}).

\subsection{Case $\kappa=1$}
At $d\neq 0$ and $\kappa=1$
substituting  another torsionless tensor field
\bq\label{l-ch}
L_g =\dfrac{1}{\sqrt{\mathrm g}\,}\,L=\dfrac{1}{\sqrt{\mathrm g}\,}\,\left(
                                 \begin{array}{cc}
                                   1 & 0 \\
                                   0 & 1
                                 \end{array}
                                \right)\,,
\eq
into the definitions (\ref{p2}), one gets  the desired solution of the equations (\ref{geom-eq})
\bq\label{p-ch}
P_g=\dfrac{1}{\sqrt{\mathrm g}}\left(
         \begin{array}{cccc}
           0 & 0 & 1 & 0 \\
           * &0 & 0 & 1 \\ \\
           * & * & 0 & -\dfrac{1}{2}\left(
          \dfrac{\partial \ln \mathrm g}{\partial \theta}p_\phi-\dfrac{\partial \ln \mathrm  g}{\partial \phi}p_\theta
                      \right) \\
           * & * & * & 0
         \end{array}
       \right)
\eq
\begin{rem}
This deformation has a similar form with the well-known relation between  the modular vector fields
\[
X_{g\mu}=X_\mu-X_{\ln g}
\]
associated with the volume forms $\mu$ and $\nu=g\mu$ \cite{kos85,wein97}.
\end{rem}

\begin{rem}
At  $d= 0$  function $\mathrm g$ (\ref{g-fun}) equals to unit  and, therefore, at this limit one gets standard  canonical Poisson bivector  $\displaystyle P=\lim_{d\to 0} P_g$.
\end{rem}

In terms of initial variables $(\g,M)$ this  Poisson bivector $P_g $ (\ref{p-ch}) has been obtained in \cite{bm01}.
\begin{prop} \cite{bm01}
Integrals of motion $H_{1,2}$ (\ref{3-int}) are in involution with respect to the Poisson bracket
associated with the Poisson  bivector $P_g $ (\ref{p-ch})
\[\{H_1,H_2\}_g =0\,,\qquad \qquad d>0,\qquad \kappa=1.\]
\end{prop}
The corresponding  volume form
\[\nu=P_g ^{-2}=- 2 \mathrm g\,\mathrm dq \mathrm dp\]
is  invariant with respect to a Hamiltonian flow associated with new  time $t_g$ defined by
\[
\dfrac{\mathrm d}{\mathrm d t_g }\,z_k=\{H_1,z_k\}_g\,,\qquad k=1,\ldots,4.
\]
We can  easily relate new and old time variables
\bq\label{ch-time}
\mathrm d t_g \simeq \sqrt{\mathrm g}\, \mathrm dt
\eq
because at $\kappa=1$ initial equations of motion are equal to
\bq\label{flow-ch}
\dfrac{\mathrm d}{\mathrm dt}\,z_k=\dfrac{\sqrt{\mathrm g}}{2}\{H_1,z_k\}_g\,.
\eq
Transformation of  time (\ref{ch-time}) has been proposed by Chaplygin in \cite{ch03}.  Namely this process
 is  to be referred to as the Chaplygin Hamiltonization, see \cite{bm01,bm08,bm08b, ekm04, bl11}.

\begin{rem}
  One of the  global invariants in Poisson geometry is a modular class. It is  an obstruction to the existence of a measure in $\mathcal M$ which is invariant under all  hamiltonian flows \cite{kos85,mar10,wein97}.  So, in fact it is a geometric obstruction to the Hamiltonization process.

  For the manifold $\mathcal M$ endowed with a Poisson bivector $P$, its modular class is an element of the first Poisson cohomology group. In Section 3 we discuss some  elements of the second Poisson cohomology group and the corresponding  Poisson bivectors $P'$  compatible with $P$, which allows us to get  variables of separation without  Hamiltonization.
\end{rem}

\subsection{Case $\kappa=-1$}

It is easy to see, that at $\kappa=-1$ the integrals of motion $H_{1,2}$ (\ref{3-int}) do not commute with respect to the Poisson brackets associated with bivector $P_g $ (\ref{p-ch})
\[\{H_1,H_2\}_g \neq 0\,,\qquad\qquad d>0\,,\qquad \kappa=-1.\]
So, we have to propose another deformation of the canonical Poisson
structure applicable to the Borisov-Mamaev-Fedorov system.

Let us try to solve our geometric equations
\bq \label{g-eq2}
(PdH_1,dH_2)\equiv\{H_1,H_2\}=0\,,\qquad [P,P]=0\,,\eq
by "brute force" method, using similar to (\ref{p-ch}) anzats
\[
P=\left(
         \begin{array}{cccc}
           0 &0 & f(\phi,\theta) & 0 \\
           * & 0 & 0 & h(\phi,\theta) \\
           * & * & 0 & u(\phi,\theta)\,p_\phi+v(\phi,\theta)\, p_\theta \\
           * & * & * & 0 \\
         \end{array}
       \right)\,.
\]
As a result we have the following
\begin{prop}
At $\kappa=-1$ the integrals of motion $H_{1,2}$ (\ref{3-int})  are in  involution with respect to the Poisson brackets associated with the Poisson bivector
\bq \label{p1-bf}
 P_\eta= \left(
 \begin{array}{cc}
 0 &{ L_\eta}_{ij} \\
 \\
 -{L_\eta}_{ij}\qquad&\displaystyle \sum_{k=1}^n\left((1+\eta)\dfrac{\partial {L_\eta}_{ki}}{\partial q_j}-
\dfrac{1} {(1+\eta)}\dfrac{\partial {L_\eta}_{kj}}{\partial q_i}\right)p_k
 \end{array}
 \right)\,,
 \eq
where
\bq\label{l-bf}
L_\eta=\dfrac{1}{\sqrt{\mathrm g}\,}\,\left(
         \begin{array}{cc}
          1 & 0 \\ \\
          0 & 1+\eta \\
         \end{array}
       \right)\,\quad\mbox{and}\quad \eta=\dfrac{2\sin^2\theta\bigl( b_3^2-(b_1+b_2)b_3+b_1b_2\bigr)}{b_3^2(d^{-1}(b_1+b_2)-2)}\,.
\eq
Here $L_\eta$ is the (1,1) tensor field with  non-zero torsion, in contrast with the  tensor field from the Turiel construction (\ref{p2}).
\end{prop}
The proof is straightforward.

Tensor field $L_\eta$  may be considered as an additional deformation  $L_g$ (\ref{l-ch}) by function $\eta$ depending only on  variable $\theta$, parameters $d$ and $b_k$, such as
 \[ \lim_{d\to0} \eta=0\,\qquad\Rightarrow\qquad \lim_{d\to0} P_\eta=P\,.
 \]
Moreover,  $\eta=0$ for the axially symmetric ball at $b_3=b_1$ or $b_3=b_2$.

 \begin{rem}
 At present we don't know any physical meaning of the function $\eta(\theta)$ and the geometric
 explanation of the deformation  (\ref{p1-bf}).  It will be interesting to understand the relations between  ${ L_\eta}$
and the theory of  Killing tensors with non-zero torsion.
 \end{rem}

 The Poisson bivector $P_\eta$ (\ref{p1-bf}) may be rewritten as follows
\[
P_\eta=\dfrac{1}{\sqrt{\mathrm g}}\left(
         \begin{array}{cccc}
           0 & 0 & 1 & 0 \\
           * &0 & 0 & (1+\eta) \\ \\
           * & * & 0 & -\dfrac{1}{2}\left(
          (1+\eta)\dfrac{\partial \ln \mathrm g}{\partial \theta}p_\phi-\dfrac{\partial \ln \mathrm g}{\partial \phi}p_\theta
                      \right) \\
           * & * & * & 0
         \end{array}
       \right)\,.
\]
The corresponding  volume form
\[\nu_\eta=P_{\eta}^{-2}=- \dfrac{2\mathrm  g}{(1+\eta)} \,\mathrm d q \mathrm d p\]
is a more complicated deformation of the invariant   volume form (\ref{rho})   introduced in  \cite{yar92}
\[ \mu= \sqrt{\mathrm g}\,\,\mathrm dq \mathrm dp\,.
\]
This new volume  form is invariant with respect to a Hamiltonian evolution associated with new time $t_\eta$ defined by
 \[
\dfrac{\mathrm d}{\mathrm d t_{\eta}}\,z_k=\{H_1,z_k\}_\eta\,,\qquad k=1,\ldots,4.
\]
Relation between the initial  and new  time variables is also more complicated  then in (\ref{ch-time}), because
at $\kappa=-1$  initial equations of motion (\ref{m-eq}) read as
\bq\label{flow-bf}
\dfrac{\mathrm d}{\mathrm dt}\,z_k=\dfrac{\sqrt{\mathrm g}}{2}\Bigl({b_1+b_2+b_3}+ w_1\Bigr)\{H_1,z_k\}_\eta-\sqrt{\mathrm g}\Bigl(1+w_2\Bigr)\{H_2,z_k\}_\eta\,.
\eq
Here
\ben
w_1&=&\dfrac{\eta}{1+\eta}\,\dfrac{(b_1+b_2)\Bigl(b_3(b_1\cos^2\phi+b_2\sin^2\phi)-b_1b_2\Bigr)}{(b_1-b_3)(b_2-b_3)}\nn\\
\label{w-bf}\\
w_2&=&\dfrac{\eta}{1+\eta}\,\dfrac{b_3(b_1\cos^2\phi+b_2\sin^2\phi)-b_1b_2}{(b_1-b_3)(b_2-b_3)}\,.\nn
\en
So, without the intermediate time transformation we did not get conformally Hamiltonian system from \cite{bm08,bm08b},
 because we consider, in fact, two different  systems with geometric point of view,  see  \cite{bl09,ekm04,bl11}. The modern theory of conformally Hamiltonian systems  may be found in \cite{mar10}.

\section{Second Poisson brackets}
\setcounter{equation}{0}
In this Section  we want to get another solution $P'$ of the equations (\ref{geom-eq},\ref{g-eq2}), which is compatible with the first solution $P$ obtained earlier, i.e.
\[
[P,P']=0\,.
\]
Compatible  bivectors  $P'$  are the 2-cocycles in the Poisson cohomology defined by  $P$ on the Poisson manifold $\mathcal M$, whereas  the Lie derivatives of $P$ along vector field $X$
\[
P'=\mathcal L_X P
\]
 are2-coboundaries.   So,   in order to get the  desired solution of (\ref{geom-eq},\ref{g-eq2}) we will use the Lie  derivatives along  the vector fields $X$ with linear in momenta entries.

 In bi-Hamiltonian geometry  equations of motion usually have the following  form
 \bq\label{bi-int}
 \dfrac{\mathrm d}{\mathrm dt}\,z_k=s_1\,\{H_1,z_k\}'+s_2\,\{H_2,z_k\}'\,,
  \eq
  where $\{.,.\}'$ is the second Poisson bracket associated with $P'$ and  $s_{1,2}$ are some functions on dynamical variables \cite{ts10d,ts11s}.

If $s_1=0$ and  $s_2=const$ we have a bi-Hamiltonian dynamical system. If  $s_1=0$ and $s_2$ is arbitrary, one gets the so-called quasi bi-Hamiltonian system. At $s_{1,2}\neq 0$ we have bi-integrable dynamical system \cite{ts07a,ts11s}.  So, the equations   of motion  (\ref{flow-bf}) for the Borisov-Mamaev-Fedorov system have the standard bi-Hamiltonian  form.

\subsection{Case $d=0$ and $\kappa=1$}
At $d=1$ we have the hamiltonian flow (\ref{m-eq}) associated with the canonical Poisson bivector $P$ (\ref{p1}).
It is easy to prove that the  integrals of motion $H_{1,2}$ (\ref{3-int})  are in bi-involution
\[
\{H_1,H_2\}=\{H_1,H_2\}'=0\,,\qquad \qquad d=0,
\]
with respect to canonical Poisson brackets associated with bivectors $P$ (\ref{p1}) and $P'$ (\ref{p2}) determined by  the following (1,1) torsionless tensor field \cite{ts11s}:
\bq\label{p2-neu}
L'=\left(
 \begin{array}{cc}
 a_1 \cos^2\phi+a_2\sin^2\phi & \dfrac{(a_1-a_2)\sin 2\phi}2\,\dfrac{\cos\theta}{\sin\theta}\\
 \\
 \dfrac{(a_1-a_2)\sin 2\phi}2\,{\cos\theta}\,{\sin\theta}\quad & a_3\sin^2\theta+(a_1\sin^2\phi+a_2\cos^2\phi)\cos^2\theta
 \end{array}
 \right)\,.
\eq
The Turiel  bivector $P'$  (\ref{p2})  may be rewritten as the Lie derivative $P'=\mathcal L_Y\,P$ of the canonical bivector $P$
 along the vector field  $Y=\sum Y^j\partial_j$ with the following entries
\bq\label{p2-ell-lie}
Y^{1,2}=0\,,\qquad \left(
                                                   \begin{array}{c}
                                                      Y^3 \\
                                                      Y^4
                                                   \end{array}
                                                 \right)=-\,{L'}^\top\left(
                                                   \begin{array}{c}
                                                     p_\phi \\
                                                      p_\theta
                                                   \end{array}
                                                 \right)\,.
\eq
Here ${L'}^\top$ stands for the transpose of the matrix ${L'}$.

\begin{rem}
The corresponding volume form $\lambda'={P'}^{-2}$ is invariant with respect to the new time defined by
\[
\dfrac{\mathrm d}{\mathrm d t'} z_k=\{H_1,z_k\}'\,,\qquad k=1,\ldots,4.
\]
It is neither bi-Hamiltonian nor quasi bi-Hamiltonian system \cite{ts11s} and, therefore,  this re\-pa\-ra\-me\-tri\-za\-tion of  time
looks like the Hamiltonization for the Borisov-Fedorov system.
\end{rem}

The eigenvalues $u,v$ of the recursion operator $N=P'P^{-1}$ are the roots of the following polynomial
\bq\label{b-pol}
B(\lambda)=(\lambda-u)(\lambda-v)=\lambda^2-\mathrm{tr}\left( L'L^{-1}\right)\,\lambda+\dfrac{\det L'}{\det L}=0\,.
\eq
Of course, in this case coordinates $u,v$ are the standard elliptic coordinates on the sphere defined by
\bq\label{uv-ell}
\dfrac{(\lambda-u)(\lambda-v)}{(\lambda-a_1)(\lambda-a_2)(\lambda-a_3)}=
\dfrac{\g_1^2}{\lambda-a_1}+\dfrac{\g_2^2}{\lambda-a_2}+\dfrac{\g_3^2}{\lambda-a_3}
\,.
\eq

\subsection{Case $d=0$ and $\kappa=-1$}
At $d=0$ and  $\kappa=-1$ the integrals of motion  $H_{1,2}$ (\ref{3-int})  are in bi-involution with respect to canonical Poisson bracket and the second bracket associated with bivector  $\widehat{P}'$  (\ref{p2}) defined by the following (1,1) tensor field
\bq\label{lp-bf}
\widehat{L}'=\left(
 \begin{array}{cc}
 c_1 \cos^2\phi+c_2\sin^2\phi & \dfrac{(c_1-c_2)\sin 2\phi}2\,\dfrac{\cos\theta}{\sin\theta}\\
 \\
 \dfrac{(c_1-c_2)\sin 2\phi}2\,{\cos\theta}\,{\sin\theta}\quad & c_3\sin^2\theta+(c_1\sin^2\phi+c_2\cos^2\phi)\cos^2\theta
 \end{array}
 \right)\,,
\eq
where  $c_i=a_i/b_i$.  The eigenvalues of the recursion operator $\widehat{N}=\widehat{P}'P^{-1}$ coincide with standard elliptic coordinates on the sphere
\bq\label{ell-bf}
\g_i=\sqrt{\dfrac{(u-c_i)(v-c_i)}{(c_j-c_i)(c_k-c_i)}},\qquad i\neq j\neq k\,,\qquad c_i=\dfrac{a_i}{b_i}\,.
\eq
The conjugated momenta  $p_u,p_v$  are defined by   standard relations
\bq\label{pell-bf}
M_i=\dfrac{1}{b_i}\,\dfrac{2\varepsilon_{ijk}\g_j\g_k(c_j-c_k)}{u-v}\Bigl((c_i-u)p_u-(c_i-v)p_v\Bigr)\,,
\eq
where $\varepsilon_{ijm}$ is a completely antisymmetric tensor.

In terms of these Darboux-Nijenhuis  variables $u,v$ and $p_u,p_v$, our Poisson bivectors look like
\[
P=\left(
    \begin{array}{cccc}
      0 & 0 & 1 & 0 \\
      0 &0 & 0 & 1 \\
      -1 & 0 & 0 & 0 \\
      0 & -1 & 0 & 0
    \end{array}
  \right)
\qquad
\widehat{P}'=\left(
    \begin{array}{cccc}
      0 & 0 & u & 0 \\
      0 &0 & 0 & v \\
      -u & 0 & 0 & 0 \\
      0 & -v & 0 & 0
    \end{array}
  \right)\,,
\]
whereas in terms of initial physical variables, first bivector $P$ reads as
 \bq\label{p1-mbf}
 P=\dfrac{1}{b_1b_2b_3}\left(
 \begin{array}{cccccc}
 0 & 0 & 0 & 0 & b_1b_3\g_3 & -b_1b_2\g_2 \\
 * & 0 & 0 & -b_2b_3\g_3 & 0 & b_2b_1\g_1 \\
 * & * & 0 & b_3b_2\g_2 & -b_3b_1\g_1 & 0 \\
 * & * & * & 0 & b_3^2\,M_3 & -b_2^2\,M_2 \\
 * & * & * & * & 0 & b_1^2\,M_1 \\
 * & * & * & * & * & 0 \\
 \end{array}
\right)\,,
 \eq

It is  evident that any functions  $f_1(u)$ and  $f_2(v)$ are variables of separation as well. So,  the following trivial point transformation
 \bq\label{triv-tr}
 u\to f_1(u)\,,\qquad v\to f_2(v)
 \eq
preserves the separability property of distribution defined by the functions  $H_{1,2}$.  Namely, according to  \cite{ts07a},  these integrals are   in involution with respect to the Poisson brackets associated with the following  bivectors
\bq\label{p-fp}
P'=\left(
    \begin{array}{cccc}
      0 & 0 & f_1(u) & 0 \\
      0 &0 & 0 & f_2(v) \\
      -f_1(u) & 0 & 0 & 0 \\
      0 & -f_2(v) & 0 & 0
    \end{array}
  \right)\,.
\eq
Of course, in terms of initial physical variables these bivectors have more complicated form. For instance, tensor field
\bq\label{lpp}
{L''}=\dfrac{1}{\zeta}\left[
\widehat{L'}+2\left(
           \begin{array}{cc}
             \rho & 0 \\
             0 & \rho
           \end{array}
         \right)
\right]
\eq
where
\ben
\zeta&=&\cos^2\theta+\dfrac{b_3(b_2+b_1)\cos^4\theta}{
\bigl(
b_1b_2+b_3(b_1\sin^2\phi+b_2\cos^2\phi)
\bigr) \sin^2\theta}\,,
\nn\\
\nn\\
\rho&=&\dfrac{\cos^2\theta}{b_1b_2+b_3(b_1\sin^2\phi+b_2\cos^2\phi)}-
\dfrac{\cos^2\theta+\cos^2\phi\sin^2\theta}{b_1(b_2+b_2)}-
\dfrac{\sin^2\phi\cos^2\theta+1}{b_2(b_1+b_3)}\,,
\nn
\en
yields bivector (\ref{p-fp}) associated with new variables of separation (\ref{triv-tr}) defined by
\ben
f_1(u)&=&-\dfrac{2\bigl(u b_1(b_2+b_3)-2\bigr)\bigl(ub_2(b_1+b_3)-2\bigr)}
{ub_1b_2(b_1+b_3)(b_2+b_3)\bigl(ub_3(b_1+b_2)-2\bigr)}\,,\nn\\
\nn\\
f_2(v)&=&-\dfrac{2\bigl(v b_1(b_2+b_3)-2\bigr)\bigl(vb_2(b_1+b_3)-2\bigr)}
{vb_1b_2(b_1+b_3)(b_2+b_3)\bigl(vb_3(b_1+b_2)-2\bigr)}\,.
\nn
\en
It is natural that the canonical transformations  (\ref{triv-tr})  preserve the first bivector and change the second bivector  $\widehat{P}'$ simultaneously with coefficients  $s_{1,2}$ in the equations of  motion (\ref{bi-int}).  Of course, some geometric properties of these equations are invariant with respect to such transformations.
\begin{prop}
At $d=0$ and $\kappa=-1$ there does not exist nontrivial linear in momenta Poisson bivector  ${{P}''}$, which is compatible with the canonical ones, such that $s_1=0$ in (\ref{bi-int}).

So, at $\kappa=-1$ the dynamical system (\ref{m-eq}) is  only bi-integrable,  whereas  at $\kappa=1$ it is bi-Hamiltonian.
\end{prop}
 By adding  equations  (\ref{bi-int}) with  $s_1=0$ and compatibility condition $[P,P'']=0$ to the initial equations  (\ref{geom-eq},\ref{g-eq2}) one gets an overdetermined system of algebro-differential equations. If the entries of $P''$ are  linear nonhomogeneous  polynomials in momenta, then the  system has  only trivial solution $P''=0$.
\begin{rem}
According to  \cite{bf95}, at $d=0$ dynamical systems with  $\kappa=\pm1$ are related to each other by the Poisson map $M\to BM$ and the trivial change of  time  \[t\to -t\,.\]
In Proposition 3 we proved that even such seemingly harmless transformation  leads to a loss of very important geometric property. Namely, after this change of time the new system becomes non bi-Hamiltonian with respect to  initial integrals of motion.
\end{rem}
\subsection{Chaplygin system,  $\kappa=1$}
According to \cite{ts11ch},   let us introduce the vector field $X=\sum X^j\partial_j$ with the following entries
\bq\label{field-ell}
X^i=0,\qquad X^{i+3}=\Bigl[\g\times \mathbf A_g (\g\times M)\Bigr]_i,\quad i=1,2,3\,.
\eq
where $\mathbf A_g$  is the following  $3\times 3$ matrix
\[
\mathbf A_g= \mathbf A +d\mathrm g(\g)\,\mathbf A\,\bigl(\g \otimes\g\bigr)\,\mathbf A\,.
\]
entering into  the angular velocity  definition (\ref{w12}).

\begin{prop}\cite{ts11ch}
The Lie derivative of  $P_g$ (\ref{p-ch}) along the vector field $X$ (\ref{field-ell}) is the desired  second  solution of the equations   (\ref{geom-eq},\ref{g-eq2}) compatible with the first solution
\bq\label{p2-ch3}
P'_g =\mathcal L_{X}\,P_g \,,
\eq
so that the  integrals of motion $H_{1,2}$ (\ref{3-int}) are in  bi-involution
\bq\label{inv-ch}
\{H_1,H_2\}_g =\{H_1,H_2\}'_g =0\,,
\eq
with respect to a pair of the corresponding compatible  Poisson brackets.
\end{prop}
In spherical coordinates this bivector looks like  a deformation of the Turiel construction (\ref{p2})
\bq \label{p2-ch4}
 P'_g= \left(
 \begin{array}{cc}
 0 &{ L'_g}_{ij} \\
 \\
 -{L'_g}_{ij}\qquad&\displaystyle \sum_{k=1}^n\left(x_{ki}\dfrac{\partial {L'_g}_{ki}}{\partial q_j}-y_{kj}\dfrac{\partial {L'_g}_{kj}}{\partial q_i}\right)p_k
 \end{array}
 \right)\,.
 \eq
 Similar to the tensor field $L_\eta$ (\ref{l-bf}) , this  (1,1) tensor field
\[
 L'_g =\sqrt{\mathrm g}\,L'-\frac{d\,\sqrt{\mathrm g}\,\sin^2\theta}{1-da_3}\left(
                                          \begin{smallmatrix}{}
                                            a_1a_2-a_3(a_1\cos^2\phi+a_2\sin^2\phi) & 0 \\ \\
                                            0 & -a_3^2+a_3(a_1\sin^2\phi+a_2\cos^2\phi)
                                          \end{smallmatrix}
                                        \right)\,.
 \]
has a non-zero torsion too.

Functions $x_{ki},y_{kj}$ depending  only on the coordinates $\phi,\theta$ can be easily restored from the relation  (\ref{p2-ch3}), which in spherical coordinates reads as
\bq\label{p2-ch-lie}
P'_g=\mathcal L_Z\,P_g\,,\qquad Z^{1,2}=0\,,\qquad \left(
                                                   \begin{array}{c}
                                                      Z^3 \\
                                                      Z^4
                                                   \end{array}
                                                 \right)=-\sqrt{\mathrm g}\,{L'_g}^\top\left(
                                                   \begin{array}{c}
                                                     p_\phi \\
                                                      p_\theta
                                                   \end{array}
                                                 \right)\,.
\eq
Here ${L'_g}^\top$ stands for the transpose of the matrix ${L'_g}$.

Using this second Poisson structure for the Chaplygin system we can rewrite the equations of motion (\ref{m-eq}) in the following form
\[
\dfrac{\mathrm d}{\mathrm dt}z_k=\dfrac{(1-d\,a_3)\sqrt{\mathrm g}}{2}\Bigl(\,\dfrac{s_1}{a_3}\,\{H_1,z_k\}'_g+\{H_2,z_k\}'_g\,\Bigr)\,,
\]
where
\[
s_1=-1-da_3+\dfrac{a_3^2\sin^2\theta-(a_3(a_1+a_2)-a_1a_2)\cos^2\theta
 -d(da_1a_2-a_1-a_2)a_3^2}
 {da_1a_2a_3-a_1a_2\cos^2\theta-a_3(a_1\cos^2\phi+a_2\sin^2\phi)\sin^2\theta}\,.
\]
The eigenvalues $ u_g, v_g$ of the  recursion operator $N_g =P'_g P^{-1}_g $ are defined by the relation
\bq\label{nhell-q}
\dfrac{(\lambda-u_g)(\lambda-v_g)}{ (\lambda-a_1)(\lambda-a_2)(\lambda-a_3)}=\mathrm g(\g)\,\left(\dfrac{\g_1^2(1-da_1)}{\lambda-a_1}+\dfrac{\g_2^2(1-da_2)}{\lambda-a_2}+
\dfrac{\g_3^2(1-da_3)}{\lambda-a_3}\right)\,,
\eq
which bears a resemblance to the usual definition (\ref{uv-ell}) of the elliptic coordinates on the sphere.
Namely these variables  have been obtained  by Chaplygin who used hamiltonization   \cite{ch03}.

As above, using the trivial point transformations (\ref{triv-tr}) we can change this second Poisson bracket and the equations of motion (\ref{bi-int}) associated with this bracket. For instance, let us consider another (1,1) tensor field
\bq
{L}''_g=\dfrac{1}{\zeta\sqrt{\mathrm g}\,}\left[
{L'}+\dfrac{1}{1-da_3}\left(
           \begin{array}{cc}
             \rho_1 & 0 \\
             0 & \rho_2
           \end{array}
         \right)
\right]
\eq
where
\ben
\zeta&=&d a_1a_2a_3+a_1a_2\cos^2\theta+a_3(a_1\cos^2\phi+a_2\sin^2\phi)\sin^2\theta\,,\nn\\
\nn\\
\rho_1&=&\Bigl(da_3-\sin^2\phi\cos^2\theta-\cos^2\phi\Bigr)a_1
+\Bigl(da_3-\cos^2\phi\sin^2\theta-1\Bigr)a_2-
a_3\sin^2\theta\,,\nn\\
\nn\\
\rho_2&=&\rho_1+d\sin^2\theta(a_1-a_3)(a_2-a_3)\,.
\nn
\en
Substituting this tensor field into the Lie derivative   (\ref{p2-ch-lie}), one gets  a new Poisson bivector ${P}''_g$  with the following properties.
\begin{prop}
At $\kappa=1$ the  initial equations of motion  (\ref{m-eq}) have the following form
\bq\label{flow-ch2}
\dfrac{\mathrm d}{\mathrm dt}\,z_k=\dfrac{\sqrt{\mathrm g}}{2}\{H_1,z_k\}_g=\dfrac{(1-da_3)\mathrm g}{2}\,\{H_2,z_k\}''_g\,,
\eq
where $\{.,.\}''_g$  is the Poisson bracket associated with bivector  $P''_g$.
\end{prop}
So,   equations of motion for the nonholonomic Chaplygin system  are conformally Hamiltonian equations with respect to  both the first bracket $\{.,.\}_g$ with  first integral of motion  $H_1$ (\ref{flow-ch}) and the second bracket  $\{.,.\}''_g$  with second integral of motion  $H_2$ (\ref{flow-ch2}).

\subsection{Borisov-Mamaev-Fedorov system, $\kappa=-1$}
As usual, at  $\kappa=-1$  there exist many linear in momenta solutions of the equations   (\ref{g-eq2}),  which are related to each other by point canonical transformations  $\lambda_i\to f_i(\lambda_i)$ (\ref{triv-tr}), where $\lambda_i$ are the Darboux-Nijenhuis coordinates, i.e. the eigenvalues of the recursion operator.

All these solutions have the form
\bq \label{p2-bf}
 P'_\eta= \left(
 \begin{array}{cc}
 0 &{ L'_\eta}_{ij} \\
 \\
 -{L'_\eta}_{ij}\qquad&\displaystyle \sum_{k=1}^n\left(x_{ki}\dfrac{\partial {L'_\eta}_{ki}}{\partial q_j}-y_{kj}\dfrac{\partial {L'_\eta}_{kj}}{\partial q_i}\right)p_k
 \end{array}
 \right)\,.
 \eq
Let us  consider only one solution  associated with  a relatively simple tensor field
\bq\label{ln-bf}
{L'_\eta}=\dfrac{1}{\sqrt{\mathrm g}}\,\left[\widehat{L}'+\left(
  \begin{array}{cc}
    \alpha & 0 \\
    0 &  (1+\eta)\alpha+\dfrac{2\eta(b_1+b_2+b_3)}{(b_1+b_2)(b_1+b_3)(b_2+b_3)} \end{array}
\right)
\right]\,,
\eq
depending on the arbitrary number $\alpha$. As above, this tensor field has a non zero torsion at  generic $\alpha$.

Functions $x_{ki},y_{kj}$ depending  only on the coordinates $\phi,\theta$ can be easily restored from the other definition
of the same bivector
\[P'_\eta=\mathcal L_Z\,P_\eta,\]
where the entries of the vector field   $Z=\sum Z^j\partial_j$ are equal to
\bq\label{p2-bf-lie}
 Z^{1,2}=0\,,\qquad \left(
                                                   \begin{array}{c}
                                                      Z^3 \\
                                                      Z^4
                                                   \end{array}
                                                 \right)=-\sqrt{\mathrm g}\,{L'_\eta}^\top\left(
                                                                                    \begin{array}{cc}
                                                                                      1 & 0 \\
                                                                                      0 & (1+\eta)^{-1} \\
                                                                                    \end{array}
                                                                                  \right)
                                                 \left(
                                                   \begin{array}{c}
                                                     p_\phi \\
                                                      p_\theta
                                                   \end{array}
                                                 \right)\,.
\eq
\begin{rem}
It is easy to see, that at $\kappa=\pm1$ entries of the Liouville vector field $Z$ can be rewritten in the common form
\[
  Z=-\left(
          \begin{array}{cc}
            0 & 0 \\
            0 & {L'_\sigma}^\top L^{-1}_\sigma \\
          \end{array}
        \right)z\,,\qquad \sigma=g, \eta\,,
\]
where  $Z=(Z^1,Z^2,Z^3,Z^4)$ is the vector of the entries vector field, whereas $z=( q_1,q_2,p_1,p_2)$ is the vector of Darboux coordinates . Geometric origin of this new construction is unclear of yet.
\end{rem}

 So, for the Borisov-Mamaev-Fedorov system we have a pair of compatible Poisson bivector on   $\mathcal M$ and, therefore,
 this manifold is bi-Hamiltonian.
\begin{prop}
At  $\kappa=-1$ the integrals of motion   $H_{1,2}$ (\ref{3-int}) are in bi-involution
 \bq\label{inv-bf}
\{H_1,H_2\}_\eta =\{H_1,H_2\}'_\eta =0\,,
\eq
with respect to the Poisson brackets associated with bivectors  $P_\eta$ and  $P'_\eta$.
\end{prop}
Using the second Poisson brackets $\{.,.\}'_\eta$, we can rewrite the initial equation of motion (\ref{m-eq}) in the standard form
(\ref{bi-int}) with relatively big coefficients $s_{1,2}$.

At $\alpha=0$ in  (\ref{ln-bf}),  the eigenvalues  $u_\eta$ and $v_\eta$  of the recursion operator  $N_\eta =P'_\eta P^{-1}_\eta$ are defined by the relation
\bq
\dfrac{(\lambda-u_\eta)(\lambda-v_\eta)}{ (\lambda-c_1)(\lambda-c_2)(\lambda-c_3-\delta)}=\dfrac{1}{\zeta(\g)}\,
\left(\dfrac{\g_1^2}{\lambda-c_1}+\dfrac{\g_2^2}{\lambda-c_2}
+ \dfrac{\beta\,\g_3^2}{\lambda-c_3-\delta}\right)\,,\label{nhell-bf}
\eq
depending on two constants
\ben
\beta&=&1-\dfrac{2d(b_1-b_3)(b_2-b_3)}{b_3(b_3-2d)(b_1+b_2)+2db_1b_2}\,,
\nn\\ \nn\\
\delta&=&\dfrac{4d}{b_3(b_3-2d)(b_1+b_2)+2db_1b_2}\,
\dfrac{b_1b_2(b_1-b_3)(b_2-b_3)}{b_3(b_1+b_2)(b_1+b_3)(b_2+b_3)}\,,
\nn\\
\en
and one function on $\theta=\arccos \g_3$
\[
\zeta(\g)=1-\dfrac{2d(b_1-b_3)(b_2-b_3)}{b_3(b_3-2d)(b_1+b_2)+2db_1b_2}\g_3^2\,.
\]
This relation is very close to  (\ref{uv-ell}) and (\ref{nhell-q}) and, of course,
at   $d=0$ coincides with the definition (\ref{ell-bf})  of the elliptic coordinates on the sphere $\mathbb S$ which are variables of separation for the corresponding Hamilton-Jacobi equation.

Now we have to compare this Darboux-Nijenhuis coordinates with the variables of separation obtained  in \cite{bmm08, bm08b}  by hamiltonization process.

\begin{prop}
 Variables of separation $\mathrm q_{1,2}$ from  \cite{bmm08, bm08b} are related with the eigenvalues $u_\eta$ and $v_\eta$
 (\ref{nhell-bf}) of the recursion operator $H_\eta$ by trivial point transformation similar to (\ref{triv-tr}).
\end{prop}
Let us reproduce the definition of variables of separation $\mathrm q_{1,2}$  from  \cite{bm08b},  see formulae  (3.2):
\bq\label{q12-bf}
\g_i=\sqrt{\dfrac{\det\mathbf I}{(J_i-d)J_j J_k\,G(\mathrm q_1,\mathrm q_2)}}\sqrt{\dfrac{(\mathrm q_1-c_i)(\mathrm q_2-c_i)}{(c_j-c_i)(c_k-c_i)}}
\eq
where
\[G(\mathrm q_1,\mathrm q_2)=\dfrac{(b_1+b_2-2d)(b_1+b_3-2d)(b_2+b_3-2d)}{(b_1+b_2)(b_1+b_3)(b_2+b_3)}\,\mathrm g\,.\]
Substituting  this definition into the equation (\ref{nhell-bf})  one gets the  desired point transformation
\[
u_\eta=F(\mathrm q_1)\,,\qquad v_\eta=F(\mathrm q_2)\,,\]
where
\[F(\mathrm q)=\dfrac{\Bigl(b_3^3+(b_1+b_2-2d)b_3^2+b_1b_2b_3+2db_1b_2\Bigr)\mathrm q -4d}{(b_1+b_3)(b_2+b_3)\bigl(d(b_1b_2 \mathrm q-2)+b_3\bigr)}\,.\]
Associated with the variables $\mathrm q_{1,2}$ tensor field $L'_\eta$  in  (\ref{p2-bf}) is  more complicated  then tensor  field  (\ref{ln-bf}) and, for brevity,  we omit  this expression.

\section{ Separation of variables}
In geometry, instead of  an additive separation of variables in the partial differential equation called the Hamilton-Jacobi equation,  we have some invariant  geometric property of the Lagrangian distribution defined by  $n$ independent functions  $H_1,\ldots,H_n$.

Namely, an $n$-tuple  $H_1,\ldots,H_n$ of functionally independent functions  defines a separable foliation on $\mathcal M$, dim$\mathcal M=n$, if there are   variables of separation $(\mathrm q_1,\dots,\mathrm q_n,\mathrm p_1,\dots,\mathrm p_n)$ and $n$ separated  relations of the form
\begin{equation}
\label{seprelint}
\Phi_i(\mathrm q_i,\mathrm p_i,H_1,\dots,H_n)=0\ ,\quad i=1,\dots,n\ ,
\qquad\mbox{with}\quad\det\left[\frac{\partial \Phi_i}{\partial H_j}\right]
\not=0\,.
\end{equation}
It simple means, the common level surfaces of $H_1,\ldots,H_n$ form foliation and every leaf of this foliation may be represented as a direct product of one-dimensional geometric objects defined by separated relations  (\ref{seprelint}). Usually
 we have a direct product of $n$ algebraic curves, because $\Phi_i$ are  polynomials in $\mathrm q_i$ and $\mathrm p_i$.

It can be easily shown  \cite{ts07a}, that condition (\ref{seprelint}) entails the involutivity  of  $H_i$ with respect to the compatible Poisson brackets
\bq
\label{poi-f}
\{\mathrm q_i,\mathrm q_j\}_f=\{\mathrm p_i,\mathrm p_j\}_f=0,\qquad \{\mathrm p_i,\mathrm q_j\}_f=\delta_{ij}\,f_j(\mathrm p_j,\mathrm q_j)\,,
\eq
depending on the arbitrary functions  $f_1,\ldots,f_n$.  In fact, this definition of separability implicitly appeared in the Lagrange  proof of the Jacobi theorem,  but both Lagrange and Jacobi used only   canonical brackets

 \bq\label{can-poib}
 \{\mathrm q_i,\mathrm q_j\}=\{\mathrm p_i,\mathrm p_j\}=0,\qquad  \{\mathrm q_i,\mathrm p_j\}=\delta_{ij}\,,
\eq
which belongs to the family (\ref{poi-f}).

 In bi-Hamiltonian geometry  eigenvalues $\mathrm q_1,\ldots,\mathrm q_n$ of the recursion operator
 are the desired coordinates of separation  or the Darboux-Nijenhuis coordinates. It is a sequence of the fact that the distribution  tangent to the foliation defined by $H_1,\ldots,H_n$ is Lagrangian with respect to the symplectic form $P^{-1}$ and invariant with respect to recursion operator $N=P'P^{-1}$. So, if we know these coordinates, then we have to explicitly  find the conjugated momenta $\mathrm p_1,\ldots,\mathrm p_n$  and the separated relations (\ref{seprelint}).

In the Chaplygin hamiltonization  method  momenta are defined by complete integrals $S$ of the Hamilton-Jacobi equation
after the corresponding change of time
\bq\label{Eq-Js} \mathrm p_j=\dfrac{\partial
}{\partial \mathrm q_j}S_j(\mathrm q_j,\alpha_1,\ldots,\alpha_n)\,.
\eq
On this step one usually gets  momenta $\mathrm p_i$   which have more complicated  Poisson brackets with coordinates   $\mathrm q_i$ (\ref{poi-f}) instead of standard canonical brackets   (\ref{can-poib}).

For instance, let us consider definition of the separated  momenta from  \cite{bm08b}, see formulae  (3.14):
\ben
M_i&=&\frac{(J_i-d)^2}{J_i^2\,b_i}\,\frac{\sqrt{(c_j-\mathrm q_1)(c_j-\mathrm q_2)}\sqrt{(c_k-\mathrm q_1)(c_k-\mathrm q_2)}}{2\sqrt{G(\mathrm q_1,\mathrm q_2)}\,(u-v)}\label{m-op}\\
\nn\\
&&\qquad\qquad\times\left(
\frac{\mathrm p_2}{(\mathrm q_1-c_j)(\mathrm q_1-c_k)}-\frac{\mathrm p_2}{(\mathrm q_2-c_j)(\mathrm q_2-c_k)}\right)\,.\nn
\en
It is easy to observe, that at  $d=0$  variables $\mathrm q_{1,2}$ (\ref{q12-bf}) coincide with the standard elliptic coordinates $u,v$ (\ref{ell-bf}) on the sphere $\mathbb S$,  but the corresponding momenta
\bq\label{p-can}
\mathrm p_1=\phi(u) p_u\,,\qquad \mathrm p_2=\phi(v) p_v
\eq
differ from the standard canonical variables  $ p_u$ and  $p_v$ (\ref{pell-bf}) on $T^*\mathbb S$.

Moreover, after substituting  $\gamma_i$ (\ref{q12-bf})  and  $M_i$ (\ref{m-op})  into $C_2=\sum b_i\g_iM_i$ one gets  $C_2\neq0$ even at $d=0$. So, we suppose that the  definition of momenta in  \cite{bm08b}  contains some misprint and, therefore, we have to  define these variables correctly.

\subsection{Chaplygin system, $\kappa=1$}
According to  \cite{ts10k,ts11ch}, we can use the following recurrence chain
\bq\label{rrel2}
\phi_1=\{u,H_k\}_g,\qquad \phi_2=\{u,\phi_1\}_g,\ldots,\quad \phi_i=\{u,\phi_{i-1}\}_g\,\,,\qquad k=1,2,
\eq
in order to calculate the desired momenta. Namely,  in our case this chain breaks down on the third step $\phi_3=0$. It means that ${H}_{1,2}$ are the second order polynomials in momenta $p_u$ and, therefore, we can define this unknown momenta in the following way
\bq\label{pkow-def1}
p_u=\dfrac{\phi_1}{\phi_2}
\eq
up to  the  canonical transformations $p_u\to p_u+ f(u)$.  Similar calculation allows us to determine the second momenta $p_v$.

At $\kappa=1$ the results obtained so far can be summarized in the following definition
\bq\label{m-ch}
M_i=\dfrac{2\varepsilon_{ijk}\g_j\g_k(a_j-a_k)\sqrt{\mathrm g}}{\mathrm{u}-\mathrm{v}}\,
\Bigl((a_i-\mathrm{u})(1-d\mathrm{u})\mathrm{p}_u-(a_i-\mathrm{v})(1-d\mathrm{v})\mathrm{p}_v\Bigr)\,.
\eq
where
\[
\mathrm g=\dfrac{(1-d\mathrm{u})(1-d\mathrm{v})}{(1-da_1)(1-da_2)(1-da_3)}\,.
\]
By adding the  expressions for  $\g_i$
\bq\label{g-ch}
\g_i=\sqrt{\dfrac{(1-da_j)(1-da_k)}{(1-d\mathrm{u})(1-d\mathrm{v})}}
\,\cdot\,\sqrt{\dfrac{(\mathrm{u}-a_i)(\mathrm{v}-a_i)}{(a_j-a_i)(a_m-a_i)}}\,
,\qquad i\neq j\neq k\,,
\eq
obtained from  (\ref{nhell-q})  to  (\ref{m-ch}) one gets the expressions of initial physical variables in terms of canonical separated variables.

By substituting (\ref{g-ch}) and    (\ref{m-ch})  into  $H_{1,2}$ (\ref{3-int}) we can easily prove that variables of separation lie on two copies of the  hyperelliptic genus 2 curve defined by the following separated relation
\bq\label{s-relCh}
 4(1-d\mathrm x)(a_1-\mathrm x)(a_2-\mathrm x)(a_3-\mathrm x)\,\mathrm y^2-\mathrm x H_2+H_1=0,\qquad \mathrm x=\mathrm{u},\mathrm{v},\quad \mathrm y=\mathrm{p}_u,\mathrm{p}_v\,.
\eq
It is easy to see, that at  $d=0$  we obtain the standard elliptic variables on  $T^*S$  and well-known separated relations for the Euler top on the sphere.

\subsection{Borisov-Mamaev-Fedorov system, $\kappa=-1$}
Let us take the coordinates of separation $\mathrm q_{1,2}$ (\ref{q12-bf}) and add momenta $\mathrm p_{1,2}$  to them, in order to obtain a complete set of the canonically conjugated variables  (\ref{can-poib})  with respect to the first Poisson bracket  (\ref{p1-bf}).

As above, we can simply  calculate these momenta  using the recurrence chain  $\phi_k$ (\ref{rrel2}) associated with the first Poisson brackets $\{.,.\}_\eta$ (\ref{p1-bf}) at $\kappa=-1$ and rewrite the obtained results  (\ref{p-can}) in the following form
\ben
M_i&=&\dfrac{2\varepsilon_{ijk}\g_j\g_k(c_j-c_k)\sqrt{\mathrm g}}{b_i(\mathrm q_1-\mathrm q_2)}
\label{m-bf}\\
\nn\\
&&\qquad\qquad\times \left(
(c_i-\mathrm q_1)\left(1-\dfrac{d(2-b_1b_2\mathrm q_1)}{b_3}\right)\mathrm p_1
-(c_i-\mathrm q_2)\left(1-\dfrac{d(2-b_1b_2\mathrm q_2)}{b_3}\right)\mathrm p_2
\right)\,,
\nn
\en
where $c_i=a_i/b_i$ and
\[
\mathrm g=\dfrac{\Bigl(1-d(b_1+b_2+b_3-2d)\mathrm q_1\Bigr)\Bigr(1-d(b_1+b_2+b_3-2d)\mathrm q_2\Bigl)}{(1-da_1)(1-da_2)(1-da_3)}+\dfrac{8d}{a_1a_2a_3}\,\mathrm q_1\mathrm q_2\,.
\]
These expressions are similar to  (\ref{m-ch}) and at $d=0$ turn into the standard definitions  (\ref{pell-bf}) of the elliptic variables on $T^8\mathbb S$, in contrast with expressions  (\ref{m-op}) from \cite{bm08b}.

By substituting $\g_i$ (\ref{q12-bf})  and  (\ref{m-bf})  into  $H_{1,2}$ (\ref{3-int}) we easily prove that variables of separation for the Borisov-Mamaev-Fedorov system lie on two   copies of the hyperelliptic genus 2 curve defined by the following separated relation,
\bq\label{s-relBf}
4\left(1-\dfrac{d(2-b_1b_2\mathrm x)}{b_3}\right)^2(\mathrm x-c_1)(\mathrm x-c_2)(\mathrm x-c_3) \mathrm y^2-\alpha\,H_2+\beta H_1=0\,,
\eq
where $\mathrm x=\mathrm q_{1,2},\quad\mathrm y=\mathrm p_{1,2}$ and
\ben
\alpha&=&db_1b_2b_3\mathrm x^2-(b_1b_2+b_1b_3+b_2b_3)\mathrm x+2
\nn\\
\nn\\
\beta&=&\left(d(b_1b_2+b_1b_3+b_2b_3)-\dfrac{(b_1+b_2)(b_1+b_3)(b_2+b_3)}{2} \right)\mathrm x
+b_1+b_2+b_3-2d\,.\nn
\en
At  $d=0$  this equation coincides with the separated equation for Hamiltonian systems on the sphere separable in elliptic variables, which lie on the elliptic curve instead of hyperelliptic at $d\neq0$.

\section{Conclusion}
Using the standard machinery of the bi-Hamitlonian geometry, we reproduce some results from  \cite{bf95, bmm08,bm08b}
obtained in framework of the Chaplygin hamiltonization method. Definitions of the Poisson bivectors for the Borisov-Mamaev-Fedorov system (\ref{p1-bf}),(\ref{p2-bf}) and of the second bivector for the Chaplygin case  (\ref{p2-ch4}) are completely new.  They can be considered as nontrivial deformations of the Turiel and Benenti constructions associated only with nonholonomic  dynamical systems.

The explicit form of the separated relations  (\ref{s-relBf}) with canonical variables of separation  for the Borisov-Mamaev-Fedorov system is also new. Because all the hyperelliptic genus 2 curves are isomorphic to each other, we could use these separated relations   (\ref{s-relCh}) and (\ref{s-relBf}) in order to get a mapping between  Chaplygin and Borisov-Mamaev-Fedorov systems. We suppose that such mapping may be extended to the case  $C_2\neq 0$, that allows us to get solutions of the equations (\ref{m-eq}) in generic case.

We would like to thank A.V. Borisov  for genuine  interest  and  helpful  discussions.

\end{document}